\documentstyle[11pt,psfig]{article}

\def\reference{\parskip 0pt\par\noindent\hangindent 0.5 truecm}

\begin{document}
%
%
\title{The spectral characteristics of the 2dFGRS--NVSS galaxies}
%


\author{C A Jackson$^1$
        \and
        D M Londish$^{2}$ 
} 
\date{}
\maketitle

{\center
$^{1}$ Research School of Astronomy \& Astrophysics, Institute
of Advanced Studies, \\ The Australian National University, Canberra,
ACT 0200, Australia \\
cjackson@mso.anu.edu.au\\[3mm]
$^2$ School of Physics, University of Sydney, NSW 2006, 
Australia\\[3mm]
}

%
\begin{abstract}

We have analysed the 2dF spectra of a sample of galaxies 
common to the 2dF galaxy redshift survey 
(2dFGRS, Colless 1999) and the NRAO VLA sky survey (NVSS, 
Condon et al. 1998).  Our sample comprises 88 galaxies  
selected by Sadler et al. (1999) from 30 2dFGRS fields
observed in 1998.
In this paper we discuss how this and future, much
larger, samples of 2dFGRS--NVSS galaxies can be interpreted
via analysis of those galaxies with strong narrow emission lines. 
Using diagnostic line ratio measurements we
confirm the majority of the eyeball classifications of Sadler
et al. (1999), although many galaxies show evidence of being `composite'
galaxies -- a mixture of AGN plus starburst components.

\end{abstract}


\bigskip

%
%

\section{Introduction}

Sources common to 2dFGRS and deep radio surveys (NVSS,FIRST,SUMSS)\footnote
{NVSS: the NRAO VLA sky survey at 1.4 GHz, Condon et al. 1998, 
FIRST: Faint images of the radio sky at 20cm, White et al. 1997 and 
SUMSS:the Sydney University Molonglo sky survey at 843 MHz, 
Bock, Large \& Sadler 1999.}
comprise large samples of local, low-to-moderate power radio sources.
The sheer number of objects (in the completed 2dFGRS, we expect $\sim$4000
galaxies in common with NVSS) and the homogeneity of the spectra will allow
us to make statistical studies of sub-populations among the radio sources,
so revealing a much clearer picture of their properties (and perhaps 
identifying rare or new classes of objects). Most importantly, the
overlap between 
2dFGRS and NVSS provides a sample from which to derive
the local radio luminosity function as 
$\sim$50\% of these 
galaxies lie at $z <$ 0.1 (Sadler et al. 1999).
A future paper will derive this local radio luminosity function
and discuss the contributions of individual populations. 
This paper reports the results of an
emission-line diagnostic analysis of the preliminary 
sample of 2dFGRS-NVSS galaxies, checking these against the intial `eyeball'
classifications of Sadler et al. (1999). 
 
Throughout this paper values of $H_0$ = 75 km s$^{-1}$ Mpc$^{-1}$
and $\Omega$=1.0 are used. \\

\newpage

\section{The sample of 2dFGRS--NVSS galaxies}

The sample of 88 2dFGRS--NVSS galaxies has been extracted from the
2dFGRS and NVSS surveys, selecting sources whose radio-optical
offset is less than 10 arcsec. Complete details of the
selection criteria, cross-matching and classification procedures are 
described by Sadler et al (1999). 

FITS files containing the 
reduced 2dF spectra for these galaxies were provided by the 2dFGRS project. \\

Each galaxy has an assigned spectral type of either `S'=starburst or `A'=AGN,
defined as follows:

\begin{description}

\item AGNs have either a pure absorption-line spectrum,
characteristic of an early-type galaxy
{\it or} an emission-line signature similar to optically-selected
Seyferts, with dominant nebular emission lines ({\it e.g.} [OIII]
and [OII]) relative to any Balmer-line emission.  As Figure 1 
shows, the majority of these objects have much lower
radio powers than usually associated with radio-loud
AGN ($\log_{10}(P_{1.4 \rm 
\thinspace GHz}) \sim 10^{21} - 10^{24}$ W Hz$^{-1}$ sr$^{-1}$). 

We can describe these AGN in terms of two sub-classes:

The absorption-line AGN have radio powers typical of
FRI radio galaxies and optical spectra typical of early-type
host galaxies. These sources show none of the characteristic
emission lines associated with starformation processes.

The emission-line AGN could be either
{\it (i)} `radio-quiet' Seyfert 2s - late-type hosts with
sub-pc scale radio core-jet structures, {\it (ii)}
emission-line FRI radio galaxies - rare but not unknown
or {\it (iii)} emission-line FRII radio galaxies
although this is unlikely as these are usually more powerful, 
$\log_{10}(P_{1.4 \rm 
\thinspace GHz}) \ge 10^{24}$ W Hz$^{-1}$ sr$^{-1}$, and are 
rare at $z <$ 0.3.

\item Starburst galaxies have emission-line signatures
similar to optically-selected star-forming galaxies, with
strong Balmer line emission relative to any other 
emission features.  Figure 2 
shows that most of these objects are of 
low radio powers, $\log_{10}(P_{1.4 \rm 
\thinspace GHz}) \le 10^{22}$ W Hz$^{-1}$ sr$^{-1}$, typical
of known starburst galaxies.

\end{description}

Our sample comprises 36 starburst galaxies 
and 52 AGNs. At the time of this analysis, FITS data was unavailable
for 3 of the 36 starburst galaxies.  Of the remaining 33,
22 have both H$\alpha$ and H$\beta$ in
emission within the 2dFGRS wavelength range ({\it i.e.}
they lie at $z <$ 0.2).   The 52 AGN can be sub-divided into 1 
broad-line Seyfert 1, 40 absorption-line 
(early-type) galaxies, 1 galaxy at $z >$ 0.2 and 
10 with both H$\alpha$ and H$\beta$ in
emission (and $z <$ 0.2). The 22 starburst galaxies and 10 AGNs with Balmer
line emission form our `narrow-line galaxy sample'. \\

Figure 2 shows the distribution of radio power with
redshift for the sample of galaxies. The starburst galaxies cluster
at low redshifts due to the flux density limit of the NVSS survey.
The AGNs generally lie at higher redshifts and radio powers. The
highest-power objects are absorption-line AGN. \\

\begin{minipage}{18cm}
\psfig{file=lumdistaba.figps,width=8.5cm,height=8cm,angle=-90}
\vspace*{-8cm}{\hspace*{9.0cm}{
\psfig{file=lumdistema.figps,width=8.5cm,height=8cm,angle=-90}}}

\vspace*{0.3in}

\psfig{file=lumdistsf.figps,width=8.5cm,height=8cm,angle=-90}

\vspace*{-7.5cm}

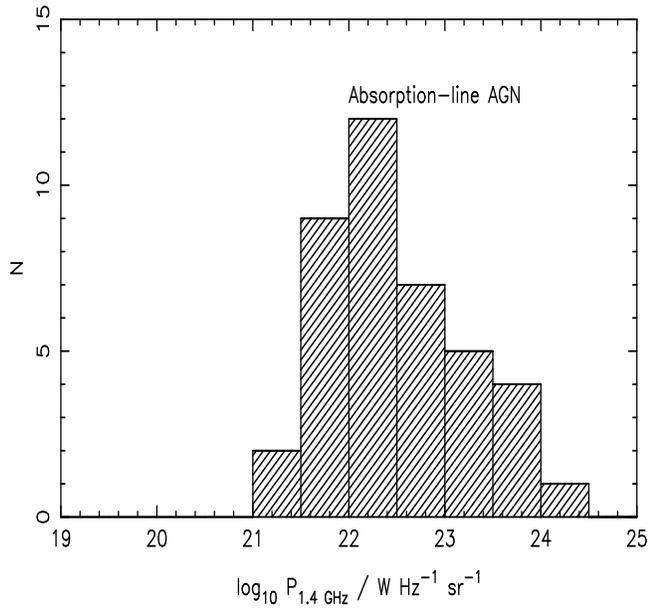
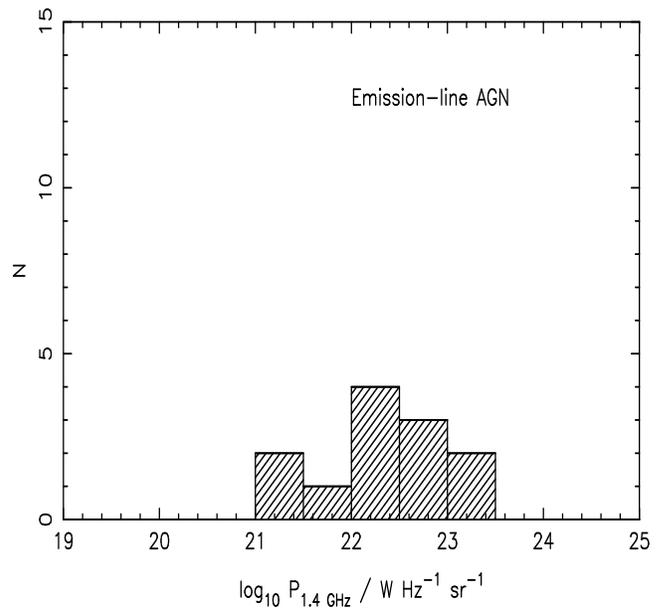
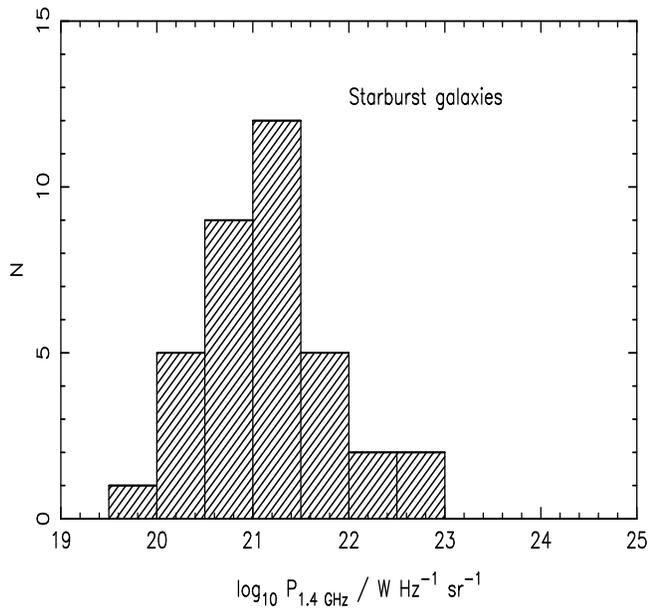
\hspace*{9.2cm}\parbox{8cm}{{\bf Figure 1.} Luminosity distributions
by type for the 88 2dFGRS--NVSS galaxies. } \\

\end{minipage}

\begin{minipage}{18cm}
\hspace*{1.0in}\psfig{file=pvz.figps,width=11cm,height=8.6cm,angle=-90}

{\bf Figure 2.} 
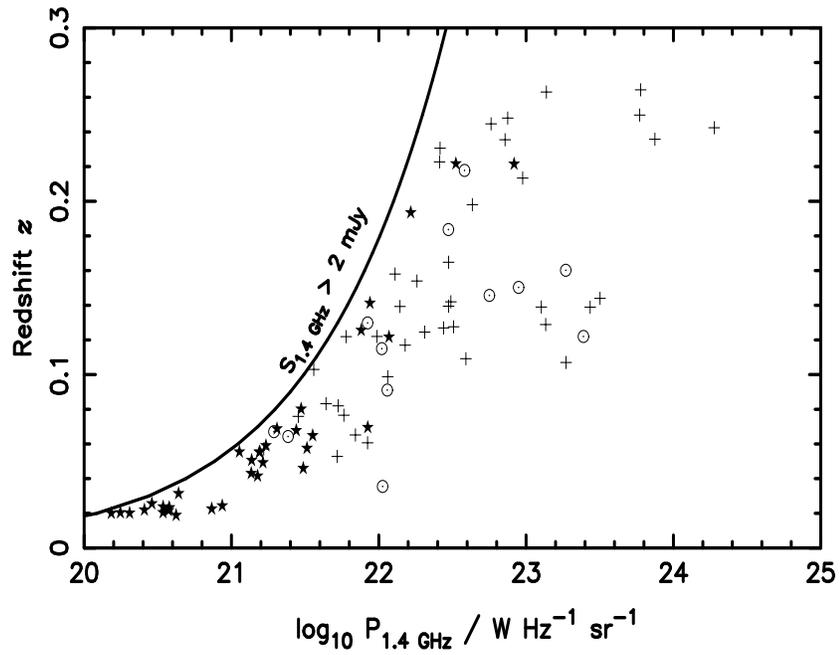
Radio power as a function of redshift for the sample of 
88 2dFGRS--NVSS galaxies. 
Symbols show the initial galaxy classification types: 
$\star$ Starburst galaxy, $\odot$ emission-line AGN
and $+$ absorption-line AGN. \\
\end{minipage}

\vspace*{6cm}

\section{Emission-line diagnostic analysis}

\vspace*{0.2in}

\begin{minipage}{18cm}
\hspace*{1.0in}\psfig{file=pvzsel.figps,width=11cm,height=8.6cm,angle=-90}

{\bf Figure 3.} 
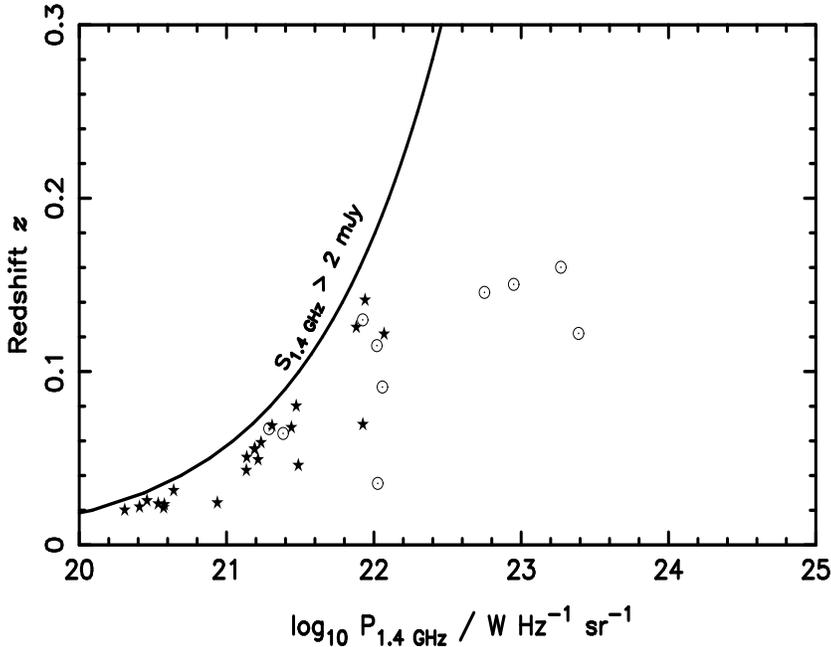
Radio power as a function of redshift for the 32 narrow-emission-line
2dFGRS--NVSS galaxies.  Symbols show the initial galaxy classification types: 
$\star$ Starburst galaxy and $\odot$ emission-line AGN. \\
\end{minipage}

We determine the emission line ratios for  
the narrow-line galaxy sample using the lines of
H$\alpha$,  H$\beta$, [OIII], [NII], [SII] and [OI].
Figure 3 shows the distribution of radio powers and redshifts 
for this sample. Comparing the distribution to Figure 1 we find
we have an unbiased (P-z) sub-set of the complete
sample. \\

Even though the 2dFGRS spectra are not flux calibrated,
the pairs of emission lines were originally
chosen to be close in wavelength. The line
ratios are therefore accurately determined even for 
our uncalibrated spectra.  The flux in each emission line is 
measured using the NOAO IRAF splot routine, fitting a Gaussian function
to each emission feature.  Diagnostic line ratios for 
[NII]$\lambda$6583/H$\alpha$,
[SII]$\lambda\lambda$6716,6731/H$\alpha$ and
[OI]$\lambda$6300/H$\alpha$ are plotted in Figure 4 and 
the data summarised in Table 1.  \\

\begin{minipage}{18cm}
\psfig{file=ratio_n2.figps,width=8cm,height=9cm,angle=-90}
\vspace*{-9cm}{\hspace*{8.5cm}{
\psfig{file=ratio_s2.figps,width=8cm,height=9cm,angle=-90}}}
\vspace*{0.3in}

\psfig{file=ratio_o1.figps,width=8cm,height=9cm,angle=-90}

\vspace*{-7.9cm}

\hspace*{8.7cm}\parbox{7cm}{{\bf Figure 4.}
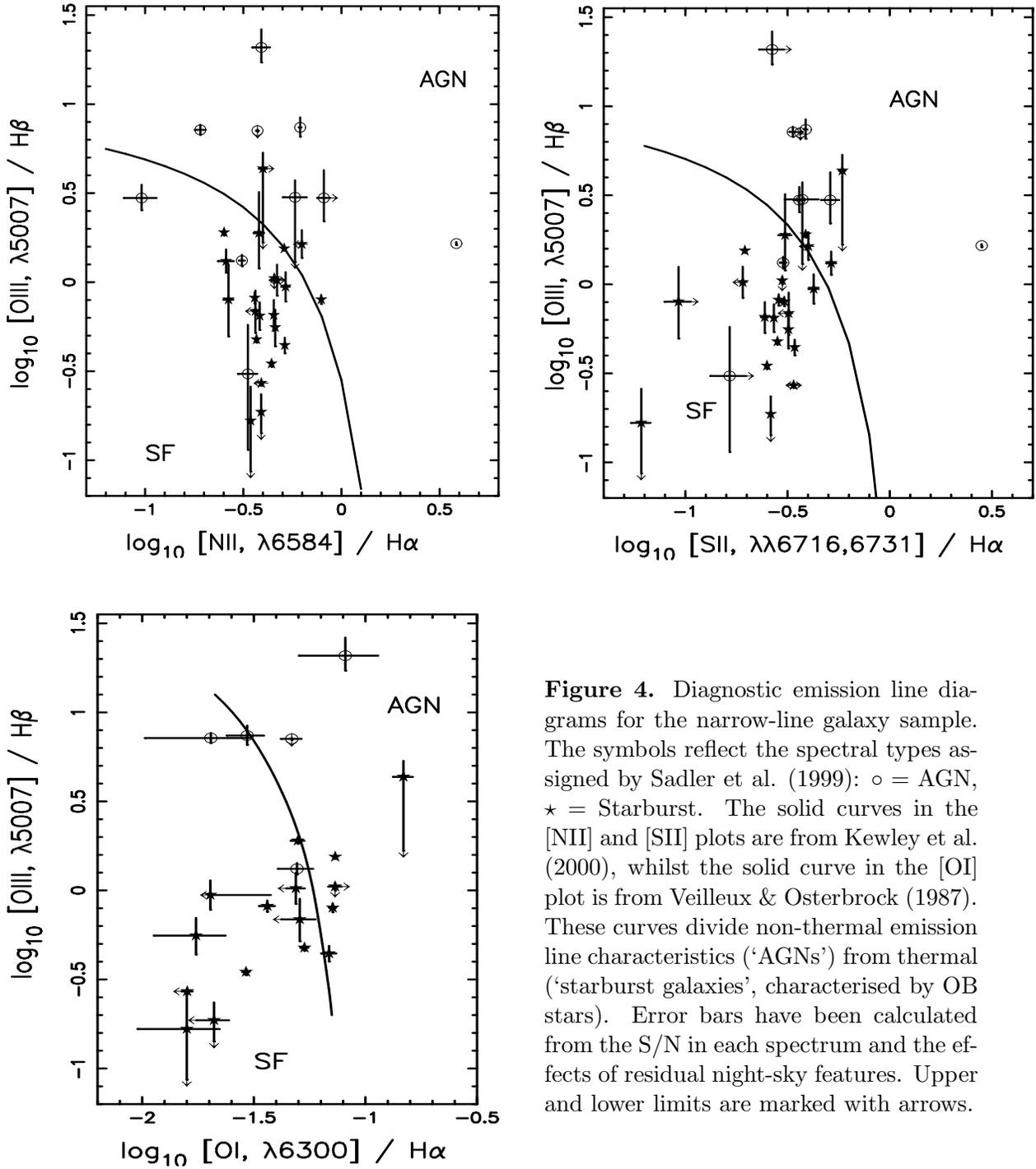
Diagnostic emission line diagrams
for the narrow-line galaxy sample. The symbols reflect the
spectral types assigned by Sadler et al. (1999):
$\circ$ = AGN, $\star$ = Starburst.
The solid curves in the [NII] and [SII]
plots are from Kewley et al. (2000), whilst the solid curve
in the [OI] plot is from Veilleux \& Osterbrock
(1987). These curves divide non-thermal
emission line characteristics (`AGNs') from
thermal (`starburst galaxies', characterised by OB stars).  
Error bars have been calculated from
the S/N in each spectrum and the effects of residual night-sky
features. Upper and lower limits are marked with arrows.} \\
\end{minipage}

\vspace*{0.5in}

\begin{table}

\begin{center}{\bf Table 1.} Emission line ratios.
\end{center}

\begin{tabular}{ccccrrrr}


2dFGRS   & & & $\log_{10}$ & 
\multicolumn{4}{c}{\rule[0.05in]{0.8in}{0.005in} Measured line ratios 
\rule[0.05in]{0.8in}{0.005in}} \\

{\small Galaxy id} & b$_{J}$ & Redshift & 
{\small P$_{1.4 \rm \thinspace GHz}\dagger$} &
\multicolumn{1}{c}{[OIII]/H$\beta$} &
\multicolumn{1}{c}{[NII]/H$\alpha$} & 
\multicolumn{1}{c}{[SII]/H$\alpha$} & 
\multicolumn{1}{c}{[OI]/H$\alpha$} \\

\hline 
\\
\multicolumn{4}{l}{\bf Initial classification SF} \\
TGS166Z108 & 17.2 & 0.0801 & 21.5 & 0.477$\pm$0.020 & 0.369$\pm$0.003 & 0.282$\pm$0.003 & 0.053$\pm$0.002  \\ 
TGS206Z164 & 14.6 & 0.0255 & 20.5 & 1.887$\pm$1.324 & 0.380$\pm$0.024 & 0.307$\pm$0.023 \\
TGS206Z015 & 17.4 & 0.0685 & 21.9 & 1.544$\pm$0.040 & 0.511$\pm$0.002 & 0.195$\pm$0.002 & 0.073$\pm$0.002  \\  
TGS232Z060 & 16.8 & 0.0590 & 21.2 & 0.443$\pm$0.046 & 0.514$\pm$0.007 & 0.342$\pm$0.007 & 0.069$\pm$0.005 \\  
TGS235Z125 & 14.8 & 0.0243 & 20.9 & $<$0.167$\pm$0.093 & 0.345$\pm$0.008 & 0.061$\pm$0.007 & 0.016$\pm$0.006  \\
TGS236Z095 & 17.6 & 0.1420 & 20.6 & $<$0.187$\pm$0.048 & 0.390$\pm$0.005 & 0.262$\pm$0.004 & $<$0.021$\pm$0.003  \\
TGS236Z091 & 17.0 & 0.0552 & 21.2 & 0.687$\pm$0.210 & $<$0.364$\pm$0.012 & $<$0.320$\pm$0.011 & $<$0.051$\pm$0.009  \\
TGS236Z194 & 15.5 & 0.0237 & 20.5 & 0.557$\pm$0.144 & 0.459$\pm$0.009 & 0.319$\pm$0.008 & 0.017$\pm$0.006  \\
TGS238Z047 & 16.5 & 0.0215 & 20.6 & 0.271$\pm$0.009 & $<$0.391$\pm$0.001 & 0.338$\pm$0.001 & 0.016$\pm$0.001  \\ 
TGS238Z180 & 17.2 & 0.1255 & 21.9 & 0.942$\pm$0.197 & 0.519$\pm$0.026 & 0.424$\pm$0.024 & $<$0.020$\pm$0.017  \\ 
TGS239Z196 & 14.8 & 0.0219 & 20.4 & 0.798$\pm$0.455 & 0.266$\pm$0.016 & $>$0.093$\pm$0.014 \\
TGS318Z156 & 18.3 & 0.0687 & 21.3 & 0.799$\pm$0.047 & 0.789$\pm$0.004 & 0.305$\pm$0.003 & 0.071$\pm$0.002  \\
TGN218Z230 & 15.9 & 0.0201 & 20.3 & 0.654$\pm$0.142 & 0.452$\pm$0.011 & 0.245$\pm$0.010 \\
TGN220Z065 & 17.3 & 0.0492 & 21.2 & $<$4.340$\pm$1.000 & $>$0.398$\pm$0.018 & 0.586$\pm$0.021 & 0.148$\pm$0.015  \\
TGN222Z132 & 16.7 & 0.0505 & 21.1 & 1.633$\pm$0.329 & $<$0.628$\pm$0.009 & $<$0.400$\pm$0.008 \\ 
TGN231Z143 & 15.5 & 0.0314 & 20.6 & 0.647$\pm$0.124 & 0.383$\pm$0.007 & 0.271$\pm$0.007 \\
TGN239Z061 & 17.5 & 0.0554 & 21.2 & 1.902$\pm$0.076 & 0.252$\pm$0.004 & 0.387$\pm$0.005 & 0.050$\pm$0.003  \\
TGN239Z082 & 18.8 & 0.1412 & 21.9 & $<$1.050$\pm$0.054 & 0.455$\pm$0.007 & 0.298$\pm$0.006 & $>$0.073$\pm$0.005  \\
TGN239Z221 & 17.4 & 0.0430 & 21.1 & 0.349$\pm$0.015 & 0.440$\pm$0.002 & 0.251$\pm$0.001 & 0.029$\pm$0.001  \\
TGN239Z172 & 16.2 & 0.0459 & 21.5 & 1.026$\pm$0.229 & 0.470$\pm$0.007 & $<$0.191$\pm$0.005 & $<$0.049$\pm$0.005  \\
TGN240Z019 & 19.3 & 0.1216 & 22.1 & 1.314$\pm$0.216 & 0.258$\pm$0.024 & 0.517$\pm$0.029 \\
TGN309Z233 & 17.9 & 0.0677 & 21.4 & 0.817$\pm$0.062 & 0.361$\pm$0.004 & 0.286$\pm$0.004 & 0.036$\pm$0.003 \\
\\
\multicolumn{4}{l}{\bf Initial classification AGN} \\
TGS207Z011 & 14.2 & 0.0355 & 22.0 &  1.649$\pm$0.035 & 3.845$\pm$0.014 & 2.817$\pm$0.011 &  0.553$\pm$0.004  \\
TGS218Z173 & 18.6 & 0.0911 & 22.1 &  7.170$\pm$0.499 & 0.191$\pm$0.012 & $>$0.335$\pm$0.013 &  0.020$\pm$0.010  \\ 
TGS220Z128 & 18.9 & 0.1298 & 21.9 & 20.817$\pm$5.495 & 0.391$\pm$0.042 & $>$0.265$\pm$0.038 &  0.081$\pm$0.033  \\ 
TGS233Z084 & 18.4 & 0.0643 & 21.4 &  2.972$\pm$0.554 & 0.096$\pm$0.019 & 0.361$\pm$0.023 \\
TGS234Z066 & 18.9 & 0.1503 & 22.9 &  $<$7.101$\pm$0.098 & 0.373$\pm$0.006 & 0.366$\pm$0.006  &  0.047$\pm$0.005  \\
TGS313Z081 & 17.7 & 0.1220 & 23.4 &  0.306$\pm$0.271 & 0.334$\pm$0.039 & $>$0.165$\pm$0.034 \\
TGS313Z100 & 17.1 & 0.1457 & 22.8 &  $<$3.000$\pm$0.741 & 0.580$\pm$0.086 & 0.374$\pm$0.075 \\
TGN231Z211 & 16.4 & 0.0671 & 21.3 &  7.402$\pm$1.041 & 0.616$\pm$0.009 & 0.388$\pm$0.008 &  0.029$\pm$0.006  \\
TGN239Z017 & 18.4 & 0.1150 & 22.0 &  2.976$\pm$1.282 & $>$0.814$\pm$0.064 & 0.511$\pm$0.054 \\
TGN238Z202 & 19.3 & 0.1602 & 23.3 &  1.323$\pm$0.103 & 0.311$\pm$0.011 & 0.301$\pm$0.011 &  0.049$\pm$0.009  \\

\hline

\end{tabular}

\vspace*{0.2in}

$^{\dagger}$ units W Hz$^{-1}$ sr$^{-1}$.

\end{table}

\newpage

The diagnostic diagrams reveal that the majority of our
narrow-line galaxy sample has [NII]$\lambda$6583/H$\alpha$,
[SII]$\lambda$6716+$\lambda$6731/H$\alpha$ 
and [OI]$\lambda$6300/H$\alpha$ 
ratios which straddle the region dividing AGN and 
starburst galaxies. This is in contrast to 
optically-selected galaxy samples, whose 
narrow-line emission characteristics clearly delineate
those with a hard ionizing radiation source 
(AGNs: right-hand side of curves in Figure 4) and those with 
hot OB stars (starburst, left-hand side of curves). 
Similar results to our line ratio distribution 
have been found for other radio-selected samples
({\it e.g.} Georgakakis et al. 1999) as well as 
IR-selected galaxy samples (Kewley et al. 2000).  

In Table 2 we show how the galaxies are classified based on the individual
emission-line ratio values. Where the errors associated with the
line ratio cross the empirical dividing line, the classification is
given as A? or S?. For the cases where the ratio lies right 
on the dividing line, the classification is given as ?. \\

We find 4 galaxies with 
different line-ratio classifications than the `eyeball' classifications
of Sadler et al. (1999):

1) TGN220Z065: Starburst galaxy, line ratio diagnostics indicate
it is an AGN. However, the H$\beta$ measurement is a lower limit so that the
the [OIII]/H$\beta$ value is over-estimated.

2) TGN222Z132: Starburst galaxy, line ratio diagnostics indicate
it is an AGN. However, the H$\alpha$ measurement is a lower limit so
that the [SII]/H$\alpha$ value is over-estimated.

3) TGN238Z202: AGN galaxy, line ratio diagnostics indicate it is
a starburst galaxy. Re-examining the 2dFGRS spectrum shows 
dominant Balmer lines characteristic of a starburst galaxy.

4) TGS313Z081: AGN galaxy, line ratio diagnostics indicate it is
a starburst galaxy.  Re-examining the 2dFGRS spectrum reveals
strong H$\alpha$, weak H$\beta$ and 
the 5577\AA- line coincident with [OIII, 5007\AA].  On balance
this probably should be classified as a starburst galaxy.

Thus the diagnostic line ratios 
confirm 30 out of 32 of the `eyeball' classifications.
However,  the frequency of ambiguous classifications from the
line-ratio diagnostics (indicated as A?, S? and ? in Table 3) 
indicates that a radio-selected sample does not clearly separate
into the starburst and AGN regions defined by
optically-selected samples.

\begin{table}

\begin{center}{\bf Table 2.} Emission-line ratio classifications 
\end{center}

\begin{tabular}{ccccccc}

 & & & & \multicolumn{3}{c}
{Line ratio classification} \\

2dFGRS   & & & $\log_{10}$ &
[NII]/ & 
[SII]/ & 
[OI]/ \\

{\small Galaxy id} & Mag b$_{J}$ & Redshift & 
{\small P$_{1.4 \rm \thinspace GHz}\dagger$} &
\multicolumn{1}{c}{H$\alpha$} & 
\multicolumn{1}{c}{H$\alpha$} & 
\multicolumn{1}{c}{H$\alpha$} \\

\hline 
\\
\multicolumn{3}{l}{\bf Initial classification SF} \\
TGS166Z108 & 17.2 & 0.0801 & 21.5 & S  & S  & S \\
TGS206Z164 & 14.6 & 0.0255 & 20.5 & S? & S? & -  \\
TGS206Z015 & 17.4 & 0.0685 & 21.9 & ?  & S  & A \\
TGS232Z060 & 16.8 & 0.0590 & 21.2 & S  & S  & ? \\
TGS235Z125 & 14.8 & 0.0243 & 20.9 & S  & S  & S \\
TGS236Z095 & 17.6 & 0.1420 & 20.6 & S  & S  & S \\
TGS236Z091 & 17.0 & 0.0552 & 21.2 & S  & S  & S \\
TGS236Z194 & 15.5 & 0.0237 & 20.5 & S  & S  & S \\
TGS238Z047 & 16.5 & 0.0215 & 20.6 & S  & S  & S \\
TGS238Z180 & 17.2 & 0.1255 & 21.9 & S  & S  & S \\ 
TGS239Z196 & 14.8 & 0.0219 & 20.4 & S  & S  & -  \\
TGS318Z156 & 18.3 & 0.0687 & 21.3 & ?  & S  & A \\
TGN218Z230 & 15.9 & 0.0201 & 20.3 & S  & S  & -  \\
TGN220Z065 & 17.3 & 0.0492 & 21.2 & A? & A  & A \\
TGN222Z132 & 16.7 & 0.0505 & 21.1 & A  & ?  & -  \\
TGN231Z143 & 15.5 & 0.0314 & 20.6 & S  & S  & -  \\
TGN239Z061 & 17.5 & 0.0554 & 21.2 & S  & ?  & ? \\
TGN239Z082 & 18.8 & 0.1412 & 21.9 & S  & S  & A \\
TGN239Z221 & 17.4 & 0.0430 & 21.1 & S  & S  & S \\
TGN239Z172 & 16.2 & 0.0459 & 21.5 & S  & S  & S \\
TGN240Z019 & 19.3 & 0.1216 & 22.1 & S  & A  & -  \\
TGN309Z233 & 17.9 & 0.0677 & 21.4 & S  & S  & S \\
\\
\multicolumn{3}{l}{\bf Initial classification AGN} \\
TGS207Z011 & 14.2 & 0.0355 & 22.0 & A  & A  & A \\
TGS218Z173 & 18.6 & 0.0911 & 22.1 & A  & A  & S \\
TGS220Z128 & 18.9 & 0.1298 & 21.9 & A  & A  & A \\
TGS233Z084 & 18.4 & 0.0643 & 21.4 & S  & A  & -  \\
TGS234Z066 & 18.9 & 0.1503 & 22.9 & A  & A  & A \\
TGS313Z081 & 17.7 & 0.1220 & 23.4 & S  & S  & -  \\
TGS313Z100 & 17.1 & 0.1457 & 22.8 & A? & A? & -  \\
TGN231Z211 & 16.4 & 0.0671 & 21.3 & A  & A  & ? \\
TGN239Z017 & 18.4 & 0.1150 & 22.0 & A  & A  & -  \\
TGN238Z202 & 19.3 & 0.1602 & 23.3 & S  & S  & S \\

\hline

\end{tabular}

\vspace*{0.2in}

$^{\dagger}$ units W Hz$^{-1}$ sr$^{-1}$.

\end{table}

\newpage

\begin{table}
\begin{center}

{\bf Table 3.} Star formation rates from IR and Radio flux densities. 

\vspace*{0.1in}

\begin{tabular}{ccrr}

       &              & \multicolumn{2}{c}{Star-formation rate} \\

2dFGRS &  $\log_{10}$ &
 \multicolumn{2}{c}{$M \ge 5M_{\odot}$ yr$^{-1}$} \\ 

galaxy id & P$_{1.4 \rm \thinspace GHz}\dagger$ &
Radio$_{1.4 \rm \thinspace GHz}$ &
IR$_{60 \rm \thinspace \mu m}$ \\

\hline
\\
TGS166Z108  & 21.5 & 9.3 & 7.1 \\ 
TGS206Z164  & 20.5 & 0.9 & 0.7 \\ 
TGS206Z015  & 21.9 & 26.4 & 77.7 \\ 
TGS232Z060  & 21.2 & 5.4 & 4.2 \\ 
TGS235Z125  & 20.9 & 2.7 & 2.1 \\ 
TGS236Z095  & 20.6 & 1.2 & 1.1 \\ 
TGS236Z091  & 21.2 & 4.9 & 2.9 \\ 
TGS236Z194  & 20.5 & 1.1 & 1.0 \\ 
TGS238Z047  & 20.6 & 1.2 & 1.3 \\ 
TGS238Z180  & 21.9 & 23.9 & 11.0 \\
TGS239Z196  & 20.4 & 0.8 & 0.5 \\ 
TGS318Z156  & 21.3 & 6.4 & 8.8 \\ 
TGN218Z230  & 20.3 & 0.6 & 0.6 \\ 
TGN220Z065  & 21.2 & 5.1 & 2.6 \\ 
TGN231Z143  & 20.6 & 1.4 & 1.0 \\ 
TGN239Z061  & 21.2 & 4.3 & 4.3 \\ 
TGN239Z172  & 21.5 & 9.7 & 7.4 \\ 

\hline

\end{tabular}
\vspace*{0.2in}

$^{\dagger}$ units W Hz$^{-1}$ sr$^{-1}$.

\end{center}

\end{table}

Further evidence as to the composite nature of the starburst galaxies can
be found by comparing the star-formation rates indicated
by the radio and IR flux densities.
Table 3 shows the flux densities and the implied star-formation
rates for massive stars ($> 5M_{\odot}$, Cram et al. (1998)), 
using 60$\mu$m flux density data compiled by Sadler et al. (1999) 
from the IRAS Point Source and Faint Source catalogues.  
Two galaxies stand out as having very discrepant implied star-formation rates:
(1) TGS206Z015 has an IR excess and
(2) TGS238Z180 has a radio excess: in both cases 
the UKST IIIaJ images show the galaxies are undergoing mergers
which is possibly enhancing or 
triggering starburst activity in these objects.

Furthermore we suggest that the majority of the narrow-line galaxies
whose line ratios are strung out between the two regions (starforming
and AGN) are composite galaxies:  The individual line ratios lie along 
mixing lines whose position on the diagnostic plots 
is determined by the admixture of AGN and starburst components
as shown in Figure 5.

\vspace*{0.1in}

\begin{minipage}{18cm}

\psfig{file=ratio_n2res.figps,width=7cm,height=8cm,angle=-90}

\vspace*{-7.9cm}

\hspace*{8.7cm}\parbox{7cm}{{\bf Figure 5.}
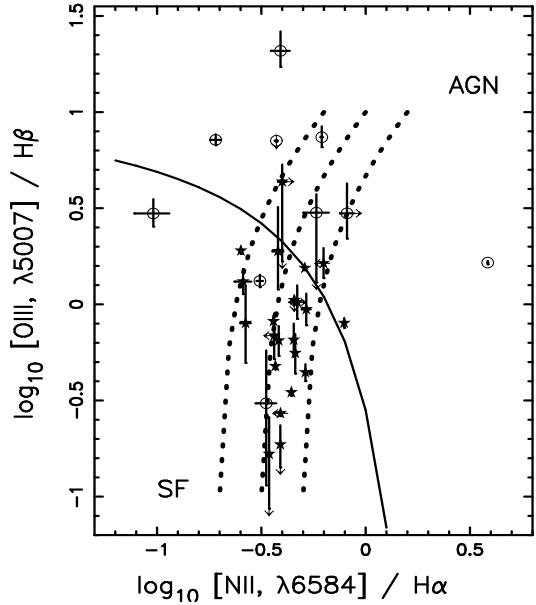
Diagnostic emission line ratios for [NII]/H$\alpha$ as shown
in Figure 4.  The solid line divides the starburst (SF) and AGN sources 
from Kewley et al. (2000). Mixing lines are shown as dotted lines, 
with increasing AGN contributions lying at higher
[OIII]/H$\beta$ values. }

\end{minipage}

\vspace*{1.6in}

\section{Summary}

Our analysis finds that this preliminary sample of narrow-emission line 
galaxies selected from 2dFGRS and NVSS comprises a mix of starburst and AGN 
types. The distribution of their emission line ratios 
suggests that a sizable fraction of 
the galaxies may be composite galaxies.  
Further work will determine the relative
contributions of the starforming and AGN components. \\

{\bf Acknowledgements} \\

We thank the 2dF Galaxy Redshift Survey team for advance access to their
data. DML thanks the Science Foundation for Physics at
the University of Sydney for a summer vacation scholarship.

%
%







\section*{References}







\reference Bock~D.C-J., Large~M.I. \& Sadler~E.M. 1999, A J, 117, 1578

\reference Colless~M.M. 1999, Phil. Trans. R. Soc. London, 357, 105

\reference Condon~J.J., Cotton~W.D., Greisen~E.W., Yin~Q.F., Perley~R.A.
\& Broderick~J.J. 1998, A J, 115, 1693

\reference Cram~L., Hopkins~A., Mobasher~B. \& Rowan-Robinson~M. 1998, ApJ,
507, 155

\reference Georgakakis~A., Mobasher~B., Cram~L., Hopkins~A., Lidman~C. \&
Rowan-Robinson~M. 1999, MNRAS, 306, 708

\reference Kewley~L., Heisler~C.A., Dopita~M.A. \& Lumsden~S. 2000, 
ApJ, submitted

\reference Sadler~E.M., McIntyre~V.J., Jackson~C.A. \& Cannon~R.D. 1999,
PASA, 16, 247

\reference Veilleux~S. \& Osterbrock~D.E. 1987, ApJS, 63, 295

\reference White~R.L., Becker~R.H., Helfand~D.J. \& Gregg~M.D. 1997, ApJ,
475, 479

\end{document}